\documentclass{article}

\usepackage{graphicx}
\usepackage{float}
\usepackage{subfigure}
\usepackage{amssymb,amsmath,mathtools,xcolor,graphicx,xspace,colortbl,ragged2e,rotating} %
\usepackage{amsfonts}  
\usepackage{amsmath}  
\usepackage{amssymb}  
\usepackage{amsthm}  
\usepackage{color}  
\usepackage{graphicx}  
\usepackage{mathtools}  
\usepackage{hyperref}
\usepackage{cite}
\usepackage{enumerate}
\usepackage{authblk}  
\graphicspath{{TanMei_graphics/}{TanMei_tcache/}{TanMei_gcache/}}
\DeclareGraphicsExtensions{.pdf,.eps,.ps,.png,.jpg,.jpeg}
\providecommand{\U}[1]{\protect\rule{.1in}{.1in}}
\allowdisplaybreaks
\newtheorem{theorem}{Theorem}

\newtheorem{definition}[theorem]{Definition}
\newtheorem{example}[theorem]{Example}

\newtheorem{notation}[theorem]{Notation}

\newtheorem{remark}[theorem]{Remark}

\begin{document}

\title{Two Variants of B\'ezout Subresultants for Several Univariate Polynomials }
\author{Weidong Wang}
\author{Jing Yang\thanks{Corresponding author: yangjing0930@gmail.com}}
\affil{HCIC--School of Mathematics and Physics, \\
Center for Applied Mathematics of Guangxi,\\
Guangxi Minzu University, Nanning 530006, China}



\date{}

\maketitle
\begin{abstract}In this paper, we develop two variants of B\'ezout subresultant formulas for several polynomials, i.e., hybrid B\'ezout subresultant polynomial and non-homogeneous B\'ezout subresultant polynomial. Rather than simply extending the variants of B\'ezout subresultant formulas developed by Diaz-Toca and Gonzalez-Vega in 2004 for two polynomials to arbitrary number of polynomials, we propose a new approach to formulating two variants of the B\'ezout-type subresultant polynomials for a set of univariate polynomials. Experimental results show that the B\'ezout-type subresultant formulas behave better than other known formulas when used to compute multi-polynomial subresultants, among which the non-homogeneous B\'ezout-type formula shows the best performance.
\end{abstract}

\section{Introduction}
Resultant and subresultant are the most important objects in resultant theory which has numerous applications (e.g., \cite{wang1998,wang2000,kapur1994,collins1991,arnon1984}).
Due to their importance,
extensive research has been carried out both in theoretical and practical aspects on resultants, subresultants and their variants\cite{sylvester1853,lascoux2003,collins1967,barnett1971greatest,terui2008,bostan2017,hong2021subresultant,cox2021,hy2021}. One of the essential topics in resultant theory is the representation of resultant and subresultant polynomials. Good representations with nice structures often bring lots of convenience for theoretical development and subsequent applications, among which determinental formulas for subresultant polynomials are a class of representations with prominent merits especially in the developments of theory and efficient algorithms. For this reason, people constructed various types of determinental formulas for subresultant polynomials since the concept was proposed, including Sylvester-type\cite{sylvester1853,li2006}, B\'ezout-type\cite{houwang2000},  Barnett-type\cite{barnett1983,diaz2002}, and so on \cite{diaz2004various}.
However, the classical subresultant polynomials are only defined for two polynomials. In \cite{hong2021subresultant}, Hong and Yang extended the concept of subresultant polynomial for two polynomials to the multi-polynomial case and gave three types of determinental formulas for the extended subresultant polynomials, i.e., Sylvester-type, B\'ezout-type and Barnett-type formulas.
These subresultant polynomials have their own interesting structures. By exploiting the hidden structures, it is expected that people may develop various algorithms for computing subresultant polynomials effectively.
It is revealed in \cite{diaz2004various} that B\'ezout matrix and its variant called hybrid B\'ezout matrix show better behavior than the Barnett matrix when used for computing the greatest common divisor of several univariate polynomials. In \cite{asadi2022subresultant}, Asadi et al. proposed a  speculative approach based on the (hybrid) B\'ezout matrix to compute the subresultant chains over rings of multivariate polynomials. For computing  subresultant  polynomials of several polynomials efficiently, it is  needed to exploit the form of known subresultants and develop new formulas from them.

In this paper,  we present two new variants of B\'ezout subresultant matrix for several univariate polynomials, i.e., hybrid B\'ezout subresultant matrix and  non-homogeneous B\'ezout subresultant matrix. It is shown that the determinants of the two matrices are equivalent to the subresultant  polynomials defined in terms of roots. The proof idea is borrowed from \cite{hong2021subresultant} and reformulated in a more friendly way.  Compared with the generalized B\'ezout subresultant polynomials for several polynomials, the two variants given in the current paper often have smaller degree. We also compare the efficiency of computing multi-polynomial subresultants with the five known subresultant formulas. It is shown that the B\'ezout formula and its two variants behave better than the Sylvester-type and Barnett-type. Among the three B\'ezout-type formulas, the non-homogeneous B\'ezout behaves best. After profiling, it is observed that the hybrid B\'ezout matrix dominates the three in forming the subresultant matrix and thus has high potentiality to be optimized when used for computing subresultants.

The paper is structured as follows. In the Section \ref{sec:pre}, we review the concepts of B\'ezout matrix and its two variants (i.e., hybrid B\'ezout matrix and non-homogeneous B\'ezout matrix) and subresultant polynomial for several polynomials. The main result of the paper is presented in the Section \ref{sec:bez} and the proof is given in Section \ref{sec:proof}. Experimental results are reported in Section \ref{sec:experiments} with further remarks.

\section{Preliminaries}\label{sec:pre}

We start with a brief introduction on the B\'ezout-type subresultant  polynomial for two univariate polynomials as well as its two variants. Then the concept  of subresultant  polynomial for several univariate polynomials is reviewed. We adopt the expression in roots of one of the given polynomials to define the subresultant
 polynomial because it is very helpful for the reasoning purpose.
Unless otherwise stated, the polynomials appearing in the rest of the paper are all univariate polynomials over the rational field, denoted by $\mathbb{Q}$, with $x$ as the variable.

\subsection{B\'ezout-type subresultant and its variants for two polynomials}

We now recall the concepts of B\'ezout matrix  and B\'ezout resultant for two polynomials as well as their two invariants including  hybrid B\'ezout matrix/resultant and non-homogeneous B\'ezout matrix/resultant.
In the rest of the subsection, we assume $A,B\in\mathbb{Q}[x]$ are with degrees $m$ and $n$, respectively, where $m\ge n$. More explicitly,
\begin{align*}
A&=a_mx^m+a_{m-1}x^{m-1}+\cdots+a_0\\
B&=b_nx^n+b_{n-1}x^{n-1}+\cdots+b_0
\end{align*}

\begin{definition}\label{def:bez}
The B\'ezout matrix  $Bez(A,B)$ of $A$ and $B$ with respect to $x$ is defined by
$${\rm{Bez}}(A,{B}): =
\left[
\begin{array}{*{20}{c}}
{c_{m-1,0}}& \cdots &{c_{m-1,m-1}}\\
 \vdots &{}& \vdots \\
c_{0,0}& \cdots &c_{0,m-1}
\end{array} \right]$$
where ${c_{i,j}}$is given by
\begin{equation}\label{eq:cij}
\frac{{{A}(x){B}(x) - {A}(y){B}(x)}}{{x - y}} = \sum\limits_{i,j = 0}^{{m} - 1} {c_{i,j}} {x^{i}}{y^{j}}
\end{equation}
The determinant of ${\rm{Bez}}(A,{B})$ is called the B\'ezout resultant of $A$ and $B$ with respect to $x$.
\end{definition}

\begin{definition}
  The hybrid B\'ezout matrix $H(A,B)$ of $A$ and $B$ with respect to $x$ is defined by
  $${{H}}({A},{B}): = \left[ {\begin{array}{*{20}{c}}
{{b_{0}}}&b_1 &\cdots &{{b_{n}}}&{}\\
{}& \ddots &\ddots&& \ddots &{}\\
{}&{}&{{b_{0}}}& b_1&\cdots &{{b_{n}}}\\
\hline
f_{1,{m}}&f_{1,{m} - 1}& \cdots &\cdots &f_{1,2}&f_{1,1}\\
 \vdots & \vdots &{}& &\vdots & \vdots \\
f_{{n},{m}}&f_{{n},{m} - 1}&\cdots & \cdots &f_{{n},2}&f_{{n},1}
\end{array}} \right]\hspace{-1em}
\begin{array}{l}
\left.
\begin{array}{l}
\\[25pt]
\end{array}
\right\}m-n\text{~rows}\\[18pt]
\left.
\begin{array}{l}
\\[25pt]
\end{array}
\right\}n\text{~rows}
\end{array}
$$
where $f_{r,j}$ is the  coefficient of the following polynomial

\begin{align*}
  k_r^{} &= ({a_{{m}}}{x^{r - 1}} +  \cdots  + {a_{m- r + 1}})({b_{{n} - r}}{x^{m- r}} +  \cdots  + {b_{0}}{x^{m- {n}}})\\
&\ \ \ \ - ({a_{m- r}}{x^{m- r}} +  \cdots  + {a_0})({b_{{n}}}{x^{r - 1}} +  \cdots  + {b_{n- r + 1}})\\
&= \sum\limits_{j = 1}^{{m}} {f_{r,j}{x^{m- j}}}
\end{align*}
in the term $x^{m-j}$ for $j=1,\ldots,m$. The determinant of ${H}(A,{B})$ is called the hybrid B\'ezout resultant of $A$ and $B$ with respect to $x$.

\end{definition}

\begin{definition}
  The non-homogeneous B\'ezout matrix $N(A,B)$ of $A$ and $B$ with respect to $x$ is defined by
$$
N({A},{B}): = \left[ {\begin{array}{*{20}{c}}
{{b_{0}}}&b_1 &\cdots &{{b_{n}}}&{}\\
{}& \ddots &\ddots&& \ddots &{}\\
{}&{}&{{b_{0}}}& b_1&\cdots &{{b_{n}}}\\
\hline
{c_{n-1,0}}&{c_{n-1,1}}& \cdots &\cdots &{c_{n-1,{m} - 2}}&{c_{n-1,{m} - 1}}\\
 \vdots & \vdots &{}&& \vdots & \vdots \\
{c_{0,0}}&{c_{0,1}}& \cdots &\cdots &{c_{0,{m} - 2}}&{c_{0,{m} - 1}}
\end{array}} \right]
\hspace{-1em}
\begin{array}{l}
\left.
\begin{array}{l}
\\[25pt]
\end{array}
\right\}m-n\text{~rows}\\[18pt]
\left.
\begin{array}{l}
\\[25pt]
\end{array}
\right\}n\text{~rows}
\end{array}$$
where ${c_{i,j}}$'s are as in \eqref{eq:cij}. The determinant of ${N}(A,{B})$ is called the non-homogeneous B\'ezout resultant of $A$ and $B$ with respect to $x$.

\end{definition}

\subsection{Subresultant in roots for several polynomials}
The following notations are needed for stating the definition of subresultant for several univariate polynomials given by Hong and Yang in \cite{hong2021subresultant}.
\begin{notation}\label{notation} \
\begin{itemize}
  \item $F = ({F_0},{F_1}, \ldots ,{F_t})\subset \mathbb{Q}[x]$;
  \item ${d_i} = \deg {F_{i}}$;
  \item ${F_0} =a_{0d_0}\prod_{j=1}^{d_0} (x-\alpha_i)$;
  \item
$\delta  = ({\delta _1},{\delta _2}, \ldots ,{\delta _t})\in \mathbb{N}^t$;
\item $\left| \delta  \right| = {\delta _1} +   \cdots  + {\delta _t} \le {d_0}$;

\item $V = \left[ {\begin{array}{*{20}{c}}
{\alpha _1^0}& \cdots &{\alpha _{{d_0}}^0}\\
 \vdots &{}& \vdots \\
{\alpha _1^{{d_0} - 1}}& \cdots &{\alpha _{d_0}^{{d_0} - 1}}
\end{array}} \right]$.
\end{itemize}
\end{notation}
With the above notations, we recall the concept of the $\delta$-th subresultant polynomial for several univariate polynomials which is defined in terms of roots of the first polynomial.

\begin{definition}\label{def:S} The generalized $\delta$-th subresultant polynomial $S_{\delta}$ of $F$ is defined by
$${S_\delta }(F): = a_{0d_0}^{\delta_0}\det M_{\delta}/\det V$$
where
\begin{itemize}
\item $M_{\delta}=\left[ {\begin{array}{*{20}{c}}
{\alpha _1^0{F_1}({\alpha _1})}& \cdots &{\alpha _{{d_0}}^0{F_1}({\alpha _{{d_0}}})}\\
 \vdots &  & \vdots \\
{\alpha _1^{{\delta _1} - 1}{F_1}({\alpha _1})}& \cdots &{\alpha _{{d_0}}^{{\delta _1} - 1}{F_1}({\alpha _{{d_0}}})}\\
\hline
 \vdots &{}& \vdots \\
 \vdots &{}& \vdots \\
\hline
{\alpha _1^0{F_t}({\alpha _1})}& \cdots &{\alpha _{{d_0}}^0{F_t}({\alpha _{{d_0}}})}\\
 \vdots &  & \vdots \\
{\alpha _1^{{\delta _t} - 1}{F_t}({\alpha _1})}& \cdots &{\alpha _{{d_0}}^{{\delta _t} - 1}{F_t}({\alpha _{{d_0}}})}\\
\hline
{\alpha _1^0(x - {\alpha _1})}& \cdots &{\alpha _{{d_0}}^0(x - {\alpha _{{d_0}}})}\\
 \vdots &  & \vdots \\
{\alpha _1^{\varepsilon  - 1}(x - {\alpha _1})}& \cdots &{\alpha _{{d_0}}^{\varepsilon  - 1}(x - {\alpha _{{d_0}}})}
\end{array}} \right]$;

  \item $\delta_0=\max(d_1+\delta_1-d_0,\ldots,d_t+\delta_t-d_0,1-|\delta|)$;

  \item $\varepsilon  = {d_0} - |\delta |$.
\end{itemize}
\end{definition}

The rational expression for $S_{\delta}$ in Definition \ref{def:S} should  be interpreted as follows, otherwise the denominator will vanish when $F$ is not squarefree.
\begin{enumerate}[(1)]
\item Treat $\alpha_1,\ldots,\alpha_n$ as indeterminates and carry out the exact division, which results in a symmetric polynomial in terms of $\alpha_1,\ldots,,\alpha_{n}$.

\item Evaluate the polynomial with $\alpha_1,\ldots,\alpha_n$  assigned the value of roots of $F$.
\end{enumerate}
Therefore, ${S_\delta }$ is essentially a polynomial in $\alpha_1,\ldots,\alpha_{d_0}$ although it is presented in the form of rational function.
Furthermore, note that ${S_\delta }$ is symmetric in $\alpha_1,\ldots,\alpha_{d_0}$ and thus it can be written as a polynomial in the coefficients of polynomials in $F$. In fact, Hong and Yang provided three representations of ${S_\delta }$ in terms of coefficients, including the Sylvester-type, the B\'ezout type and the Barnett-type subresultants. In particular,  the explicit formula for the B\'ezout-type subresultant for $F$ is presented below. The construction of the B\'ezout-type subresultant inspires us with a promising approach to construct the hybrid B\'ezout-type and non-homogeneous B\'ezout-type subresultants.

\begin{theorem}\label{thm:bezsres}
Assume $d_0=\max_{0\le i\le t}d_i$ and $\delta\ne(0,\ldots,0)$. Let
\[ {{\rm Bez}_\delta }(F): = {\left[ {\begin{array}{*{20}{l}}{{R_1}}&{{R_2}}& \cdots &{{R_t}}&{{X_{\delta ,{d_0}}}}
\end{array}} \right]^T}
\]
where \begin{itemize}
\item
$R_i$ consists of the first $\delta_i$ columns of $ {\rm Bez}(F_0,F_i)$, and

\item
$\ X_{\delta,d_0}=%
\begin{array}
[c]{l}%
\begin{bmatrix}
x &  &  \\
-1 & \ddots &  \\
& \ddots & x & \\
&  & -1 &\\
&  &  & \\
&  &  &
\end{bmatrix}\hspace{-.8em}
\left.
\begin{array}
[c]{c}%
\\
\\
\\
\\
\\
\\
\
\end{array}
\right\} d_{0}~\text{rows}\\[-8pt]
~\underbrace{\hspace{7em}}_{d_{0}-\left|  \delta\right|
~\text{columns}~}%
\end{array}
$

\end{itemize}
 Then we have
$${S_\delta }= a_{0d_{0}}^{\delta_{0}-\left|  \delta\right|  }\det
{{\rm Bez}_\delta }(F).
$$
\end{theorem}

\section{Main Results}\label{sec:bez}
In this section, we propose a new approach to construct the hybrid B\'ezout matrix and  non-homogeneous B\'ezout subresultant matrix for a set of univariate polynomials, which is different from the way developed by Diaz-Toca and Gonzalez-Vega in \cite{diaz2004various}. We will show that the determinants of the two matrices are identical with  the subresultant  polynomial of the given polynomial set.

In \cite{hong2021subresultant}, Hong and Yang proposed a method for constructing the B\'ezout subresultant matrix for several polynomials from the B\'ezout matrices  ${\rm Bez}(F_0,F_1),\ldots,$ ${\rm Bez}(F_0,F_t)$. Following the similar idea, we  construct the  hybrid B\'ezout subresultant matrix and non-homogeneous B\'ezout subresultant matrix for more than two univariate polynomials below.
For stating the main result, we assume
$F_i=a_{id_i}x^{d_i}+\cdots+a_{i0}$ for $i=0,1,\ldots,t$ where $d_0=\max_{0\le i\le t}d_i$ and
$${\rm Bez}(F_0,F_i)
 =
\left[
\begin{array}{*{20}{c}}
{c_{d_{0}-1,0}^{(i)}}& \cdots &{c_{d_{0}-1,{d_{0}} - 1}^{(i)}}\\
 \vdots &{}& \vdots \\
c_{0,0}^{(i)}& \cdots &c_{0,d_{0}-1}^{(i)}
\end{array} \right]$$

\begin{definition}\label{def:hb}
Given $F=(F_0,F_1,\ldots,F_t)$ where $F_i=\sum_{j=0}^{d_i}a_{ij}x^j$,  the generalized $\delta$-th hybrid B\'ezout subresultant matrix $H_{\delta}$ of $F$ is defined by
$${H_\delta }(F): = {\left[ {\begin{array}{*{20}{l}}{{R_1}}&{{R_2}}& \cdots &{{R_t}}&{{X_{\delta ,{d_0}}}}
\end{array}} \right]^T}$$
where $R_i$ is the transpose of the submatrix of $H(F_0,F_i)$ obtained by selecting its first $\delta_i$ rows, that it,
$$
\setlength{\arraycolsep}{4pt}
{R_i} = {\left[ {\begin{array}{*{20}{c}}
a_{i0}&\cdots &{{a_{i{d_i}}}}&{}&{}\\
&\ddots &{}& \ddots &{}\\
&&{{a_{i0}}}& \cdots &{{a_{i{d_i}}}}\\
\hline
{f_{1,{d_0}}^{(i)}}& \cdots & \cdots &{f_{1,2}^{(i)}}&{f_{1,1}^{(i)}}\\
 \vdots & & &\vdots & \vdots \\
{f_{{\delta _i} + {d_i} - {d_0},{d_0}}^{(i)}}& \cdots& \cdots  &{f_{{\delta _i} + {d_i} - {d_0},2}^{(i)}}&{f_{{\delta _i} + {d_i} - {d_0},1}^{(i)}}
\end{array}} \right]^T}
\hspace{-2em}\begin{array}{*{20}{l}}
{\left. {\begin{array}{*{20}{c}}
{}\\
{}\\
{}
\end{array}} \right\}}{\min ({\delta _i},{d_0} - {d_i})} \text{~rows}\\[15pt]
{\left. {\begin{array}{*{20}{c}}
{}\\
{}\\
{}
\end{array}} \right\}}{\max (0,{\delta _i} + {d_i} - {d_0})}\text{~rows}
\end{array}$$
and
$f^{(i)}_{r,j}$ is the  coefficient of the following polynomial

\begin{align}
  k_r^{(i)} &= ({a_{{0d_0}}}{x^{r - 1}} +  \cdots  + {a_{0d_0- r + 1}})({a_{{id_i} - r}}{x^{d_0- r}} +  \cdots  + {a_{i0}}{x^{d_0- {d_i}}})\notag\\
&\ \ \ \ - ({a_{0d_0- r}}{x^{d_0- r}} +  \cdots  + {a_{00}})({a_{{id_i}}}{x^{r - 1}} +  \cdots  + {a_{id_i- r + 1}})\label{eq:k_r}\\
&= \sum\limits_{j = 1}^{{d_{0}}} {f_{r,j}^{(i)}{x^{d_0- j}}}\notag
\end{align}
in the term $x^{d_{0}-j}$ for $j=1,\ldots,d_0$.
\end{definition}

\begin{definition}\label{def:nh}
Given $F=(F_0,F_1,\ldots,F_t)$ where $F_i=\sum_{j=0}^{d_i}a_{ij}x^j$,  the generalized $\delta$-th non-homogenous B\'ezout subresultant matrix $N_\delta$  of $F$ is defined by
$$N_\delta(F): = {\left[ {\begin{array}{*{20}{l}}
{{R_1}}&{{R_2}}& \cdots &{{R_t}}&{{X_{\delta ,{d_0}}}}
\end{array}} \right]^T}$$
where $R_i$ is the transpose of the submatrix of $N(F_0,F_i)$ obtained by selecting its firt $\delta_i$ rows, that it,
$${R_i} =
\setlength{\arraycolsep}{5pt}{\left[ {\begin{array}{*{20}{c}}
{{a_{i0}}}& \cdots &{{a_{i{d_i}}}}&{}&{}\\
{}& \ddots &{}& ~~~~~~~~~~~\ddots &{}\\
{}&{}&{{a_{i0}}}& \cdots &{{a_{i{d_i}}}}\\
\hline
{c_{d_i-1,0}^{(i)}}& \cdots&\cdots &{c_{d_i-1,{d_0} - 2}^{(i)}}&{c_{d_i-1,{d_0} - 1}^{(i)}}\\
 \vdots & & &\vdots & \vdots \\
{c_{d_{0}-\delta_i,0}^{(i)}}& \cdots&\cdots &{c_{d_{0}-\delta_i,{d_0} - 2}^{(i)}}&{c_{d_{0}-\delta_i,{d_0} - 1}^{(i)}}
\end{array}} \right]^T}\hspace{-2em}\begin{array}{*{20}{l}}
{\left. {\begin{array}{*{20}{c}}
{}\\
{}\\
{}
\end{array}} \right\}}{\min ({\delta _i},{d_0} - {d_i})\text{~rows}}\\[15pt]
{\left. {\begin{array}{*{20}{c}}
{}\\
{}\\
{}
\end{array}} \right\}}{\max (0,{\delta _i} + {d_i} - {d_0})\text{~rows}}
\end{array}$$
\end{definition}

\begin{remark}
The matrices $H_\delta(F)$ and $N_\delta(F)$ can be viewed as a generalization of the subresultant matrix developed by Li in \cite{li2006} for the Sylvester-type subresultant polynomial of two univariate polynomials.
\end{remark}

\begin{theorem}[Main result]\label{the:bez}
We have
\begin{description}
\item[(1)] ${S_\delta }(F) = c \cdot \det H_\delta (F)$,

\item[(2)] ${S_\delta }(F) =c \cdot \det N_\delta (F)$,
\end{description}
where
$c= a_{{0d_0}}^{\delta_0-\sum_{i = 1}^t {\max (0,{\delta _i} + {d_i} - {d_0})} }$.
\end{theorem}

\begin{remark}\ 
\begin{enumerate}[(1)]
\item 
The difference between the construction of B\'ezout-type subresultant variants in this paper and that in \cite{diaz2004various} is that we select rows to formulate the subresultant matrices while the latter selects columns. In the two-polynomial case, both approaches produce the same subresultant polynomials. 
\item Note that $\max (0,{\delta _i} + {d_i} - {d_0})\le\delta_i $ and thus $\sum_{i = 1}^t {\max (0,{\delta _i} + {d_i} - {d_0})}\le|\delta|$. Therefore,  when compared with the generalized B\'ezout subresultant polynomials developed in \cite{hong2021subresultant},
the two invariants of B\'ezout-type subresultant polynomials developed in the current paper often have smaller degrees.
\end{enumerate}
\end{remark}

\begin{example}
Consider~$F= ({F_0},{F_1},{F_2})$ where
\begin{align*}
  F_0 &=  {{a_{05}}{x^5}+{a_{04}}{x^4} + a_{03}}{x^3} + {a_{02}}{x^2} + {a_{01}}x + {a_{00}},\\
  F_1 &=  {{a_{14}}{x^4} + a_{13}}{x^3} + {a_{12}}{x^2} + {a_{11}}x + {a_{10}},\\
  F_2 &= {{a_{24}}{x^4} + a_{23}}{x^3} + {a_{22}}{x^2} + {a_{21}}x + {a_{20}}.
\end{align*}
and~$a_{05}a_{14}a_{24}\ne0$. Let~$\delta=(2,2)$. By Definitions
\ref{def:hb} and \ref{def:nh},
\begin{align*}
{H}_\delta(F)&= \left[ \begin {array}{ccccc} a_{{10}}&-a_{{00}}a_{{14}}&a_{{20}}&
-a_{{00}}a_{{24}}&x\\ \noalign{\medskip}a_{{11}}&-a_{{01}}a_{{14}
}+a_{{05}}a_{{10}}&a_{{21}}&-a_{{01}}a_{{24}}+a_{{05}}a_{{20}}&
-1\\ \noalign{\medskip}a_{{12}}&-a_{{02}}a_{{14}}+a_{{05}}a_{{11}
}&a_{{22}}&-a_{{02}}a_{{24}}+a_{{05}}a_{{21}}&0
\\ \noalign{\medskip}a_{{13}}&-a_{{03}}a_{{14}}+a_{{05}}a_{{12}}&
a_{{23}}&-a_{{03}}a_{{24}}+a_{{05}}a_{{22}}&0
\\ \noalign{\medskip}a_{{14}}&-a_{{04}}a_{{14}}+a_{{13}}a_{{05}}&
a_{{24}}&-a_{{04}}a_{{24}}+a_{{23}}a_{{05}}&0\end {array}
 \right]^T,\\
{N}_\delta(F)&=\left[ \begin {array}{ccccc} a_{{10}}&-a_{{00}}a_{{14}}+a_{{04}}a
_{{10}}&a_{{20}}&-a_{{00}}a_{{24}}+a_{{04}}a_{{20}}&x
\\ \noalign{\medskip}a_{{11}}&-a_{{01}}a_{{14}}+a_{{04}}a_{{11}}+
a_{{05}}a_{{10}}&a_{{21}}&-a_{{01}}a_{{24}}+a_{{04}}a_{{21}}+a_
{{05}}a_{{20}}&-1\\ \noalign{\medskip}a_{{12}}&-a_{{02}}a_{{14}}+
a_{{04}}a_{{12}}+a_{{05}}a_{{11}}&a_{{22}}&-a_{{02}}a_{{24}}+a_
{{04}}a_{{22}}+a_{{05}}a_{{21}}&0\\ \noalign{\medskip}a_{{13}}&-a
_{{03}}a_{{14}}+a_{{04}}a_{{13}}+a_{{05}}a_{{12}}&a_{{23}}&-a_{
{03}}a_{{24}}+a_{{04}}a_{{23}}+a_{{05}}a_{{22}}&0
\\ \noalign{\medskip}a_{{14}}&a_{{13}}a_{{05}}&a_{{24}}&a_{{23}}a
_{{05}}&0\end {array} \right]^T.
\end{align*}
Further calculation yields
\begin{align*}
\delta_0&=\max(\delta_1+d_1-d_0,\tilde{\delta_2+d_2-d_0}, 1-(\delta_1+\delta_2))=1,\\
c &={a_{05}^{\delta_0-(\max(0,\delta_1+d_1-d_0)+\max(0,\delta_2+d_2-d_0))}}={a_{05}^{-1}}.
\end{align*}
By Theorem \ref{the:bez}, we have
\[
{S_\delta }(F) = {a_{05}^{-1}} \cdot \det {H}_\delta(F)={a_{05}^{-1}}\cdot \det{N}_\delta(F).\]

If one computes ${S_\delta }(F)$ with B\'ezout subresultant matrix of $F$, then by Theorem \ref{thm:bezsres}, 
$${S_\delta }(F) ={a_{05}^{-3}}\cdot\det {\rm Bez}_\delta(F), $$
which indicates that $\det {\rm Bez}_\delta(F)$ has a higher degree than
$\det {H}_\delta(F)$ and $\det {N}_\delta(F)$.
\end{example}

\section{Proof}\label{sec:proof}
In this section, we show the proof of Theorem \ref{the:bez}.
\subsection{Proof of Theorem \ref{the:bez}-(1)}
\begin{proof}
By Definition \ref{def:S}, we only need to show that
  $${S_\delta }(F) \cdot \det V =  c \cdot \det (H_\delta (F) \cdot V)$$
Next we will keep simplifying the  determinant of $H_\delta (F) \cdot V$.

Consider the product $H_\delta (F) \cdot V$:
$$H_\delta (F) \cdot V=\begin{bmatrix}
R_1^T\\\vdots\\[3pt]R_t^T\\[3pt]X_{\delta ,{d_0}}^T
\end{bmatrix}\cdot V=\begin{bmatrix}
R_1^TV\\\vdots\\[3pt]R_t^T V\\[3pt]X_{\delta ,{d_0}} ^TV
\end{bmatrix}$$
where
\[R_i= {\left[ {\begin{array}{*{20}{c}}
a_{i0}&\cdots &{{a_{i{d_i}}}}&{}&{}\\
&\ddots &{}& \ddots &{}\\
&&{{a_{i0}}}& \cdots &{{a_{i{d_i}}}}\\
\hline
{f_{1,{d_0}}^{(i)}}& \cdots & \cdots &{f_{1,2}^{(i)}}&{f_{1,1}^{(i)}}\\
 \vdots & & &\vdots & \vdots \\
{f_{{\delta _i} + {d_i} - {d_0},{d_0}}^{(i)}}& \cdots& \cdots  &{f_{{\delta _i} + {d_i} - {d_0},2}^{(i)}}&{f_{{\delta _i} + {d_i} - {d_0},1}^{(i)}}
\end{array}} \right]^T}
\hspace{-1.4em}\begin{array}{*{20}{l}}
{\left. {\begin{array}{*{20}{c}}
{}\\
{}\\
{}
\end{array}} \right\}}{\min ({\delta _i},{d_0} - {d_i})} \text{~rows}\\[15pt]
{\left. {\begin{array}{*{20}{c}}
{}\\
{}\\
{}
\end{array}} \right\}}{\max (0,{\delta _i} + {d_i} - {d_0})}\text{~rows}
\end{array}\]
Meanwhile, we partition the denominator $M_{\delta}$ of $S_\delta (F)$, into $t+1$ parts, that is,
\begin{equation}\label{eq:partition_of_M}
M_\delta (F)=\begin{bmatrix}
M_1\\
\vdots\\
M_t\\
X_{\varepsilon}
\end{bmatrix}
\end{equation}
where
\begin{align*}
M_i=&
\begin{bmatrix}
{\alpha _1^0{F_i}({\alpha _1})}& \cdots &{\alpha _{{d_0}}^0{F_i}({\alpha _{{d_0}}})}\\
 \vdots &  & \vdots \\
{\alpha _1^{{\delta _1} - 1}{F_i}({\alpha _1})}& \cdots &{\alpha _{{d_0}}^{{\delta _1} - 1}{F_i}({\alpha _{{d_0}}})}
\end{bmatrix}\\
X_{\varepsilon}=&
\begin{bmatrix}
{\alpha _1^0(x - {\alpha _1})}& \cdots &{\alpha _{{d_0}}^0(x - {\alpha _{{d_0}}})}\\
 \vdots &  & \vdots \\
{\alpha _1^{\varepsilon  - 1}(x - {\alpha _1})}& \cdots &{\alpha _{{d_0}}^{\varepsilon  - 1}(x - {\alpha _{{d_0}}})}
\end{bmatrix}
\end{align*}
We will show that there exists a $\delta_i\times \delta_i$ matrix $T_i$ such that $T_iM_i=R_i^TV$ and
$X_{\delta,d_0}^TV=X_{\varepsilon}$.

\begin{enumerate}
\item Show that $T_iM_i=R_i^TV$.

Note that
{\scriptsize
\[
R_i^TV
\setlength{\arraycolsep}{0.5pt}
= \left[ {\begin{array}{*{20}{c}}
{\alpha _1^0(a_{i0}^{}\alpha _1^0 +  \cdots  + a_{i{d_i}}^{}\alpha _1^{{d_i}})}& \cdots &{\alpha _{{d_0}}^0(a_{i0}^{}\alpha _{{d_0}}^0 +  \cdots  + a_{i{d_i}}^{}\alpha _{{d_0}}^{{d_i}})}\\
 \vdots &{}& \vdots \\
{\alpha _1^{{d_0} - {d_1} - 1}(a_{i0}^{}\alpha _1^0 +  \cdots  + a_{i{d_i}}^{}\alpha _1^{{d_i}})}& \cdots &{\alpha _{{d_0}}^{{d_0} - {d_1} - 1}(a_{i0}^{}\alpha _{{d_0}}^0 +  \cdots  + a_{i{d_i}}^{}\alpha _{{d_0}}^{{d_1}})}\\
\hline
{f_{1,{d_0}}^{(i)}\alpha _1^0 +  \cdots  + f_{1,1}^{(i)}\alpha _1^{{d_0} - 1}}& \cdots &{f_{1,{d_0}}^{(i)}\alpha _{{d_0}}^0 +  \cdots  + f_{1,1}^{(i)}\alpha _{{d_0}}^{{d_0} - 1}}\\
 \vdots &{}& \vdots \\
{f_{{\delta _i} + {d_i} - {d_0},{d_0}}^{(i)}\alpha _1^0 +  \cdots  + f_{{\delta _i} + {d_i} - {d_0},1}^{(i)}\alpha _1^{{d_0} - 1}}& \cdots &{f_{{\delta _i} + {d_i} - {d_0},{d_0}}^{(1)}\alpha _{{d_0}}^0 +  \cdots  + f_{{\delta _i} + {d_i} - {d_0},1}^{(1)}\alpha _{{d_0}}^{{d_0} - 1}}
\end{array}} \right]
\]
}

From the above matrix, we have the following observations:
\begin{itemize}
  \item $a_{i0}^{}\alpha _j^0 +  \cdots  + a_{i{d_i}}^{}\alpha _j^{{d_i}} = {F_i}({\alpha _j})$;
  \item $f_{r,{d_0}}^{(i)}\alpha _j^0 +  \cdots  + f_{r,1}^{(i)}\alpha _j^{{d_0} - 1} = k_r^{(i)}({\alpha _j})$.
\end{itemize}
Thus we have
$${R_i^TV} = \left[ {\begin{array}{*{20}{c}}
{\alpha _1^0{F_i}({\alpha _1})}& \cdots &{\alpha _{{d_0}}^0{F_i}({\alpha _{{d_0}}})}\\
 \vdots &{}& \vdots \\
{\alpha _1^{{d_0} - {d_i} - 1}{F_i}({\alpha _1})}& \cdots &{\alpha _{{d_0}}^{{d_0} - {d_i} - 1}{F_i}({\alpha _{{d_0}}})}\\
\hline
{k_1^{(i)}({\alpha _1})}& \cdots &{k_1^{(i)}({\alpha _{{d_0}}})}\\
 \vdots &{}& \vdots \\
{k_{{\delta _i} + {d_i} - {d_0}}^{(i)}({\alpha _1})}& \cdots &{k_{{\delta _i} + {d_i} - {d_0}}^{(i)}({\alpha _{{d_0}}})}
\end{array}} \right]$$
Recall \eqref{eq:k_r}. Plugging $x=\alpha_i$ into it, we obtain:
\begin{align*}
    k_r^{(i)}({\alpha _j}) &= ({a_{{0d_0}}}\alpha _j^{r - 1} +  \cdots  + {a_{{0d_0} - r + 1}})(a_{i{d_i} - r}^{}\alpha _j^{{d_0} - r} +  \cdots  + a_{i0}^{}\alpha _j^{{d_0} - {d_i}}) \\
     & \  \ \ \  -({a_{{0d_0} - r}}\alpha _j^{{d_0} - r} +  \cdots  + {a_{00}})(a_{i{d_i}}^{}\alpha _j^{r - 1} +  \cdots  + a_{i{d_i} - r + 1}^{}) \\
    &= ({a_{{0d_0}}}\alpha _j^{r - 1} +  \cdots  + {a_{{0d_0} - r + 1}})(a_{i{d_i} - r}^{}\alpha _j^{{d_0} - r} +  \cdots  + a_{i0}^{}\alpha _j^{{d_0} - {d_i}})\\
    &\ \ \  \ + ({a_{{0d_0}}}\alpha _j^{{d_0}} +  \cdots  + {a_{{0d_0} - r + 1}}\alpha _j^{{d_0} - r + 1})(a_{i{d_i}}^{}\alpha _j^{r - 1} +  \cdots  + a_{i{d_i} - r + 1}^{})\\
    &=({a_{{0d_0}}}\alpha _j^{r - 1} +  \cdots  + {a_{{0d_0} - r + 1}})\cdot(a_{i0}^{}\alpha _j^{{d_0}} +  \cdots  + a_{i0}^{}\alpha _j^{{d_0} - {d_i}})\\
    &=\alpha _j^{{d_0} - {d_i}}{F_i}({\alpha _j})({a_{{0d_0}}}\alpha _j^{r - 1} +  \cdots  + {a_{{0d_0} - r + 1}})
\end{align*}
which immediately yields that
$$\setlength{\arraycolsep}{2pt}
{R_i^TV} = \left[ \begin{array}{*{20}{c}}
 {\alpha _1^0{F_i}({\alpha _1})}& \cdots &{\alpha _{{d_0}}^0{F_i}({\alpha _{{d_0}}})}\\
  \vdots &{}& \vdots \\
 {\alpha _1^{{d_0} - {d_i} - 1}{F_i}({\alpha _1})}& \cdots &{\alpha _{{d_0}}^{{d_0} - {d_i} - 1}{F_i}({\alpha _{{d_0}}})}\\
 \hline
\alpha_1^{{d_0} - {d_i}}{F_i}({\alpha_1})G_{0}(\alpha_1)& \cdots &
 \alpha_{{d_0}}^{{d_0} - {d_i}}{F_i}({\alpha_{{d_0}}})G_{0}(\alpha_{d_0})\\
  \vdots && \vdots \\
{\alpha _1^{{d_0} - {d_i}}{F_i}({\alpha _1})G_{\delta_i+d_i-d_0-1}(\alpha_1)}& \cdots &
{\alpha_{{d_0}}^{{d_0} - {d_i}}{F_i}({\alpha _{{d_0}}})G_{\delta_i+d_i-d_0-1}(\alpha_{d_0})}
\end{array} \right]
$$
where
$G_{r}(\alpha_j)={a_{{0d_0}}}\alpha _j^{r - 1} +  \cdots  + {a_{{0d_0} - r + 1}}$. We continue to simplify the lower part of $R_i^TV$ (which has $\max(0,\delta_i+d_i-d_0)$ rows) with a series of row operations.

Observing that
$$\begin{bmatrix}
G_{0}(\alpha_j)\\
G_{1}(\alpha_j)\\
\vdots\\
G_{\delta_i+d_i-d_0-1}(\alpha_j)
\end{bmatrix}=\begin{bmatrix}
a_{0d_0}\\
a_{0d_0-1}&a_{0d_0}\\
\vdots&\vdots&\ddots\\
a_{02d_0-\delta_i-d_i+1}&\cdot&\cdots&a_{{0d_0}}
\end{bmatrix}
\begin{bmatrix}
\alpha_j^{0}\\\alpha_j^1\\\vdots\\\alpha_j^{\delta_i+d_i-d_0-1}
\end{bmatrix}$$
we immediately have
\begin{align*}&
\begin{bmatrix}
\alpha_1^{{d_0} - {d_i}}{F_i}({\alpha_1})G_{i,0}(\alpha_1)& \cdots &
 \alpha_{{d_0}}^{{d_0} - {d_i}}{F_i}({\alpha_{{d_0}}})G_{i,0}(\alpha_{d_0})\\
  \vdots && \vdots \\
{\alpha _1^{{d_0} - {d_i}}{F_i}({\alpha _1})G_{i,\delta_i+d_i-d_0-1}(\alpha_1)}& \cdots &
{\alpha_{{d_0}}^{{d_0} - {d_i}}{F_i}({\alpha _{{d_0}}})G_{i,\delta_i+d_i-d_0-1}(\alpha_{d_0})}
\end{bmatrix}\\
=&\begin{bmatrix}
a_{0d_0}\\
a_{0d_0-1}&a_{0d_0}\\
\vdots&\vdots&\ddots\\
a_{02d_0-\delta_i-d_i+1}&\cdot&\cdots&a_{{0d_0}}
\end{bmatrix}\begin{bmatrix}
{\alpha _1^{{d_0} - {d_i}}{F_i}({\alpha _1})}& \cdots &{\alpha _{{d_0}}^{{d_0} - {d_i}}{F_i}({\alpha _{{d_0}}})}\\
 \vdots &{}& \vdots \\
{\alpha _1^{{\delta _i} - 1}{F_i}({\alpha _1})}& \cdots &{\alpha _{{d_0}}^{{\delta _i} - 1}{F_i}({\alpha _{{d_0}}})}
\end{bmatrix}
\end{align*}

Hence, let
$$\tilde{T}_i=\begin{bmatrix}
a_{0d_0}\\
a_{0d_0-1}&a_{0d_0}\\
\vdots&\vdots&\ddots\\
a_{02d_0-\delta_i-d_i+1}&\cdot&\cdots&a_{{0d_0}}
\end{bmatrix}$$
which has the order ${\max (0,{\delta _i} + {d_i} - {d_0})}$.
Then
\[{R_i^TV} =\begin{bmatrix}
I_i&\\
&\tilde{T}_i\end{bmatrix}
\left[ {\begin{array}{*{20}{c}}
{\alpha _1^0{F_i}({\alpha _1})}& \cdots &{\alpha _{{d_0}}^0{F_i}({\alpha _{{d_0}}})}\\
 \vdots &{}& \vdots \\
{\alpha _1^{{d_0} - {d_i} - 1}{F_i}({\alpha _1})}& \cdots &{\alpha _{{d_0}}^{{d_0} - {d_i} - 1}{F_i}({\alpha _{{d_0}}})}\\
\hline
{\alpha _1^{{d_0} - {d_i}}{F_i}({\alpha _1})}& \cdots &{\alpha _{{d_0}}^{{d_0} - {d_i}}{F_i}({\alpha _{{d_0}}})}\\
 \vdots &{}& \vdots \\
{\alpha _1^{{\delta _i} - 1}{F_i}({\alpha _1})}& \cdots &{\alpha _{{d_0}}^{{\delta _i} - 1}{F_i}({\alpha _{{d_0}}})}
\end{array}} \right]
\]
where $I_{i}$ is of order ${\min ({\delta _i},{d_0} - {d_i})}$. Let
$_{}T_{i}=\begin{bmatrix}
I_i&\\
&\tilde{T}_i\end{bmatrix}$.
Then $T_{i}$ is  of order $\delta_i$ and  ${R_i^TV} =T_iM_i$.

\item Show that $X_{\delta,d_0}^TV=X_{\varepsilon}$.

It is easy to be verified by carrying out the following matrix product:
\begin{align*}
X_{\delta,d_0}^TV&=
\begin{bmatrix}
x&-1&&&&\\
&\ddots&\ddots&&&\\
&&x&-1&&
\end{bmatrix}_{\varepsilon\times d_0}
\begin{bmatrix}
{\alpha _1^0}& \cdots &{\alpha _{{d_0}}^0}\\
 \vdots &{}& \vdots \\
{\alpha _1^{{d_0} - 1}}& \cdots &{\alpha _{d_{0}}^{{d_0} - 1}}
\end{bmatrix}\\
&=\begin{bmatrix}
\alpha_1^0(x-\alpha_1)&\cdots&\alpha_{d_0}^0(x-\alpha_{d_0})\\
\vdots&&\vdots\\
\alpha_1^{\varepsilon-1}(x-\alpha_1)&\cdots&\alpha_{d_0}^{\varepsilon-1}(x-\alpha_{d_0})
\end{bmatrix}X_{\varepsilon}
\end{align*}
\end{enumerate}

To sum up, we have
$$\begin{bmatrix}
T_1&&&\\
&\ddots&&\\
&&T_t&\\
&&&I
\end{bmatrix}\begin{bmatrix}
M_1\\\vdots\\[3pt]M_t\\[3pt]X_{\varepsilon}
\end{bmatrix}=\begin{bmatrix}
R_1^TV\\\vdots\\[3pt]R_t^T V\\[3pt]X_{\delta ,{d_0}} V
\end{bmatrix}=H_\delta (F) \cdot V
$$
Finally, taking determinants on the left and right sides, we obtain the following:
\[
\prod_{i=1}^t\det T_i\cdot\det M_{\delta}=\det H_\delta (F) \cdot \det V\]
where
\[
\det T_i=\det \begin{bmatrix}
I_i&\\
&\tilde{T}_i\end{bmatrix}=\det \tilde{T}_i
\]
Recall that $\tilde{T}_i$ is of order $\max(0,\delta_i+d_i-d_0)$ and is a lower-triangular matrix with diagonal entries to be $a_{0d_0}$. Thus
\[\det \tilde{T}_i=a_{0d_0}^{\max(0,\delta_i+d_i-d_0)}\]
which yields $\det T_i=a_{0d_0}^{\max(0,\delta_i+d_i-d_0)}$.
Then it is easy to derive that
\begin{align*}
S_{\delta}(F)&=a_{0d_0}^{\delta_0}\cdot\det M_{\delta}/\det V\\
&=a_{0d_0}^{\delta_0}\det H_\delta (F) \bigg/ {\prod_{i=1}^t\det T_i}
\\
&=a_{0d_0}^{\delta_0-\sum_{i=1}^t\max(0,\delta_i+d_i-d_0)}\det H_\delta (F)
\end{align*}
\end{proof}

\subsection{Proof of Theorem \ref{the:bez}-(2)}
\begin{proof}
By Definition \ref{def:S}, we only need to show that
  $${S_\delta }(F) \cdot \det V =  c \cdot \det (N_\delta (F) \cdot V)$$
Next we will keep simplifying the  determinant of $N_\delta (F) \cdot V$.

Consider the product $N_\delta (F) \cdot V$. We have
$$N_\delta (F) \cdot V=\begin{bmatrix}
R_1^T\\[3pt]\vdots\\[3pt]R_t^T\\[3pt]X_{\delta ,{d_0}}^T
\end{bmatrix}\cdot V=\begin{bmatrix}
R_1^TV\\[3pt]\vdots\\[3pt]R_t^T V\\[3pt]X_{\delta ,{d_0}} ^TV
\end{bmatrix}$$
where
\[
{R_i} =
\setlength{\arraycolsep}{5pt}{\left[ {\begin{array}{*{20}{c}}
{{a_{i0}}}& \cdots &{{a_{i{d_i}}}}&{}&{}\\
{}& \ddots &{}& ~~~~~~~~~~~\ddots &{}\\
{}&{}&{{a_{i0}}}& \cdots &{{a_{i{d_i}}}}\\
\hline
{c_{d_i-1,0}^{(i)}}& \cdots&\cdots &{c_{d_i-1,{d_0} - 2}^{(i)}}&{c_{d_i-1,{d_0} - 1}^{(i)}}\\
 \vdots & & &\vdots & \vdots \\
{c_{d_{0}-\delta_i,0}^{(i)}}& \cdots&\cdots &{c_{d_{0}-\delta_i,{d_0} - 2}^{(i)}}&{c_{d_{0}-\delta_i,{d_0} - 1}^{(i)}}
\end{array}} \right]^T}\hspace{-2.3em}\begin{array}{*{20}{l}}
{\left. {\begin{array}{*{20}{c}}
{}\\
{}\\
{}
\end{array}} \right\}}{\min ({\delta _i},{d_0} - {d_i})\text{~rows}}\\[15pt]
{\left. {\begin{array}{*{20}{c}}
{}\\
{}\\
{}
\end{array}} \right\}}{\max (0,{\delta _i} + {d_i} - {d_0})\text{~rows}}
\end{array}\]
As done in \eqref{eq:partition_of_M}, we partition the denominator $M_{\delta}$ of $S_\delta (F)$, into $t+1$ parts, denoted by $M_1,\ldots,M_t,X_{\varepsilon}$.
By the proof of Theorem \ref{the:bez}-1, $X_{\delta,d_0}^TV=X_{\varepsilon}$. It remains to show $R_i^TV=T_iM_i$ for some $\delta_i\times \delta_i$ matrix $T_i$.

Note that
\[
R_i^TV
\setlength{\arraycolsep}{4pt}
= \left[ {\begin{array}{*{20}{c}}
{\alpha _1^0{F_i}({\alpha _1})}& \cdots &{\alpha _{{d_0}}^0{F_i}({\alpha _{{d_0}}})}\\
 \vdots &{}& \vdots \\
{\alpha _1^{{d_0} - {d_i} - 1}{F_i}({\alpha _1})}& \cdots &{\alpha _{{d_0}}^{{d_0} - {d_i} - 1}{F_i}({\alpha _{{d_0}}})}\\\hline
{C_{{d_i}-1}^{(i)} \cdot {{\bar \alpha }_1}}& \cdots &{C_{{d_i}-1}^{(i)} \cdot {{\bar \alpha }_{{d_0}}}}\\
 \vdots &{}& \vdots \\
{C_{{d_0-\delta _i}}^{(i)} \cdot {{\bar \alpha }_1}}& \cdots &{C_{{d_0-\delta_i}}^{(i)} \cdot {{\bar \alpha }_{{d_0}}}}
\end{array}} \right]
\]
where
$$C_k^{(i)} \cdot \;{\bar \alpha _j} = \left[ {\begin{array}{*{20}{c}}
{c_{k ,0}^{(i)}}&{c_{k ,1}^{(i)}}& \cdots &{c_{k,{d_0} - 1}^{(i)}}
\end{array}} \right] \cdot {\begin{bmatrix}
{\alpha _j^0}\\[3pt]{\alpha _j^1}\\[3pt] \vdots \\[3pt]{\alpha _j^{{d_0} - 1}}
\end{bmatrix}}$$
Now we partition $R_i^TV$ into two blocks, i.e.,
\[R_i^TV=\begin{bmatrix}U_1\\[3pt]U_2\end{bmatrix}\]
with
\begin{align*}
U_1&=\begin{bmatrix}
{\alpha _1^0{F_i}({\alpha _1})}& \cdots &{\alpha _{{d_0}}^0{F_i}({\alpha _{{d_0}}})}\\
 \vdots &{}& \vdots \\
{\alpha _1^{{d_0} - {d_i} - 1}{F_i}({\alpha _1})}& \cdots &{\alpha _{{d_0}}^{{d_0} - {d_i} - 1}{F_i}({\alpha _{{d_0}}})}
\end{bmatrix}\\
U_2&=\begin{bmatrix}
{C_{{d_i}-1}^{(i)} \cdot {{\bar \alpha }_1}}& \cdots &{C_{{d_i}-1}^{(i)} \cdot {{\bar \alpha }_{{d_0}}}}\\
 \vdots &{}& \vdots \\
{C_{{d_0-\delta _i}}^{(i)} \cdot {{\bar \alpha }_1}}& \cdots &{C_{{d_0-\delta_i}}^{(i)} \cdot {{\bar \alpha }_{{d_0}}}}
\end{bmatrix}
\end{align*}
We continue to simplify $U_2$ (which has $\max(0,\delta_i-d_0+d_i)$ rows) with a series of row operations.

Recall \cite[Lemma 35]{hong2021subresultant} which states that
$$C_k^{(i)} \cdot \;{\bar \alpha _j} = {a_{{0d_0}}}{F_i}({\alpha _j}){( - 1)^{{d_0} - k-1}}e_{{d_0} - k-1}^{(j)}$$
where
$e_{\ell}^{(j)}$ denotes the $\ell$-th elementary symmetric function on  ${\alpha _1},{\alpha _2}, \ldots ,{\alpha _{j - 1}},$ ${\alpha _{j + 1}}, \ldots ,{\alpha _{{d_0}}}$.
Substituting the above equation into  $U_{2}$ and factoring $a_{0d_0}$ out, we have
$$U_{2}= a_{0d_0}\left[
\begin{array}{*{20}{c}}
F_i(\alpha _1)( - 1)^{d_0 - d_i}e_{d_0 - d_i}^{(1)}& \cdots &F_i(\alpha_{d_0})( - 1)^{d_0 - d_i}e_{d_0 - d_i}^{(d_0)}\\
 \vdots &{}& \vdots \\
{F_i}({\alpha _1})( - 1)^{\delta _i - 1}e_{\delta _i - 1}^{(1)}& \cdots &F_i(\alpha _{d_0})( - 1)^{\delta _i - 1}e_{\delta _i - 1}^{(d_0)}
\end{array} \right]
$$
By \cite[Lemma 36]{hong2021subresultant},
$$e_j^{(i)} = \sum\limits_{k = 0}^j {{{( - 1)}^k}} {e_{j - k}}\alpha _i^k=
[(-1)^0e_j~(-1)^1e_{j-1}~\cdots~(-1)^je_0~0~\cdots~0]\cdot \begin{bmatrix}
{\alpha _i^0}\\{\alpha _i^1}\\ \vdots \\{\alpha _i^{{d_0} - 1}}
\end{bmatrix} $$
where $e_\ell$ is the $\ell$-th elementary symmetric polynomial on $\alpha_1,\ldots,\alpha_{d_0}$
with the convention $e_0^{(i)}:=0$.
Denote $[(-1)^0e_j~(-1)^1e_{j-1}~\cdots~(-1)^je_0~0~\cdots~0]$ with $\bar{e}_j$. Then $e_j^{(i)}=\bar{e}_j\bar{\alpha}_i$ and thus
\begin{align*}
U_2&=a_{0d_0}
\begin{bmatrix}
{F_i}({\alpha _1}){{( - 1)}^{{d_0} - {d_i}}}\bar{e}_{{d_0} - {d_i}}\bar{\alpha}_1& \cdots &{{F_i}({\alpha _{{d_0}}}){{( - 1)}^{{d_0} - {d_i}}}}\bar{e}_{{d_0} - {d_i}}\bar{\alpha}_{d_0}\\
 \vdots &{}& \vdots \\
{F_i}({\alpha _1}){{( - 1)}^{{\delta _i} - 1}}\bar{e}_{\delta_i-1}\bar{\alpha}_1& \cdots &{{F_i}({\alpha _{{d_0}}}){{( - 1)}^{{\delta _i} - 1}}}\bar{e}_{\delta_i-1}\bar{\alpha}_{d_0}
\end{bmatrix}\\
&=a_{{0d_0}}\begin{bmatrix}
{{( - 1)}^{{d_0} - {d_i}}}\bar{e}_{{d_0} - {d_i}}\\
\vdots\\
{{( - 1)}^{{\delta _i} - 1}}\bar{e}_{\delta_i-1}
\end{bmatrix}
\begin{bmatrix}
\bar{\alpha}_1&\cdots&\bar{\alpha}_{d_0}
\end{bmatrix}
\begin{bmatrix}
{F_i}({\alpha _1})\\
&\ddots&\\
&&{F_i}({\alpha _{d_0}})\end{bmatrix}
\end{align*}
Noting that
the last $d_0-\delta_i$ columns of $\bar{e}_{{d_0} - {d_i}},\ldots,\bar{e}_{\delta_i-1}$ are all zeros, we truncate these columns and denote the resulting vectors with $\tilde{e}_{{d_0} - {d_i}},\ldots,\tilde{e}_{\delta_i-1}$. With the the last $d_0-\delta_i$ rows of $\begin{bmatrix}
\bar{\alpha}_1&\cdots&\bar{\alpha}_{d_0}
\end{bmatrix}$ cancelled by these zero columns,   we obtain
$$U_2=\tilde{T}_i
\begin{bmatrix}
{\alpha}_1^0&\cdots&{\alpha}_{d_0}^0\\
\vdots&&\vdots\\
{\alpha}_1^{\delta_i-1}&\cdots&{\alpha}_{d_0}^{\delta_i-1}
\end{bmatrix}
\begin{bmatrix}
{F_i}({\alpha _1})\\
&\ddots&\\
&&{F_i}({\alpha _{d_0}})\end{bmatrix}
$$
where
\[\tilde{T}_i=a_{{0d_0}}\begin{bmatrix}
{{( - 1)}^{{d_0} - {d_i}}}\tilde{e}_{{d_0} - {d_i}}\\
\vdots\\
{{( - 1)}^{{\delta _i} - 1}}\tilde{e}_{\delta_i-1}
\end{bmatrix}
\]
It is easy to see that $\tilde{T}_i$ is of order $\max(0, \delta_i-d_0+d_i)\times\delta_i $.

On the other hand, it is observed that
\[U_1=\begin{bmatrix}
I_i&0
\end{bmatrix}
\begin{bmatrix}
{\alpha}_1^0&\cdots&{\alpha}_{d_0}^0\\
\vdots&&\vdots\\
{\alpha}_1^{\delta_i-1}&\cdots&{\alpha}_{d_0}^{\delta_i-1}
\end{bmatrix}
\begin{bmatrix}
{F_i}({\alpha _1})\\
&\ddots&\\
&&{F_i}({\alpha _{d_0}})\end{bmatrix}\]
where the order of $I_i$ is $\min(\delta_i, d_0-d_i)$.
We construct
\[T_i=\left[\begin{array}{c}
I_{i}\qquad 0\\\hline
\tilde{T}_i
\end{array}\right]\]
and it follows that
\[R_i^TV=\begin{bmatrix}U_1\\[3pt]U_2\end{bmatrix}=T_i\begin{bmatrix}
{\alpha}_1^0&\cdots&{\alpha}_{d_0}^0\\
\vdots&&\vdots\\
{\alpha}_1^{\delta_i-1}&\cdots&{\alpha}_{d_0}^{\delta_i-1}
\end{bmatrix}
\begin{bmatrix}
{F_i}({\alpha _1})\\
&\ddots&\\
&&{F_i}({\alpha _{d_0}})\end{bmatrix}=T_iM_i\]

Finally assembling $R_i^TV$ together, we achieve the following:
\[
N_\delta (F) \cdot V=\begin{bmatrix}
R_1^TV\\[3pt]\vdots\\[3pt]R_t^T V\\[3pt]X_{\delta ,{d_0}} ^TV
\end{bmatrix}=\begin{bmatrix}
T_1M_1\\
\vdots\\
T_tM_t\\
X_{\varepsilon}
\end{bmatrix}=\begin{bmatrix}
T_1\\
&\ddots\\
&&T_t\\
&&&I_{\varepsilon}
\end{bmatrix}\begin{bmatrix}
M_1\\
\vdots\\
M_t\\
X_{\varepsilon}
\end{bmatrix}
\]
where $I_{\varepsilon}$ is the identity matrix of order $\varepsilon$.
Taking determinant on both sides yields
\[\det N_\delta (F) \cdot \det V=\prod_{i=1}^t\det T_i\cdot \det M_{\delta}\]
Further calculation derives
\[\det T_i=\det \left[\begin{array}{c}
I\qquad 0\\\hline
\tilde{T}_i
\end{array}\right]=a_{0d_0}^{\sum_{i=1}^t\max(0,\delta_i-d_0+d_i)}\]
which immediately implies
$$S_{\delta}(F)=a_{0d_0}^{\delta_0}\det M_{\delta}/\det V=a_{0d_0}^{\delta_0}\cdot\det N_\delta (F)\cdot\frac{1}{\prod_{i=1}^t\det T_i}=c\cdot\det N_\delta (F) $$
where
\[c=a_{0d_0}^{\delta_0-\sum_{i=1}^t\max(0,\delta_i-d_0+d_i)}\]
\end{proof}

\section{Experimental Results}\label{sec:experiments}

In this section, we run a collection of examples to examine the efficiency for computing the subresultant polynomials with various subresultant formulas. The involved formulas includes the Sylvester type, the Barnett type, and the B\'ezout type as well as its two variants developed in the current paper. These examples are run on a PC equipped with the Intel Core i7-10710U processor and a 16.0G RAM. In particular, the comparison is carried out from two aspects. One is the time cost for computing different  subresultant polynomials with the same polynomial set as  $\delta$ changes. The other is the time cost charged by each stage in the computation of  multi-polynomial subresultant polynomials.

\begin{figure}[H]
\centering
\subfigure[The degrees of polynomials are $(15,12,9)$]{\includegraphics[height=4.5cm]{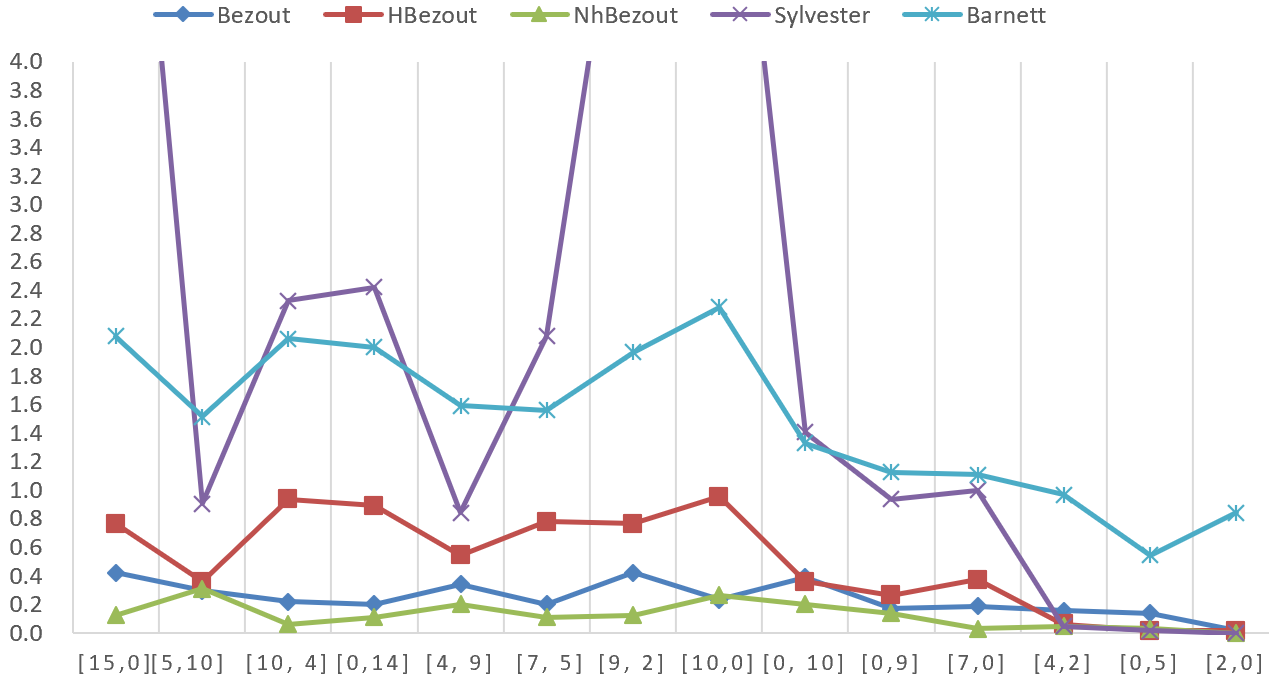}}
\\ %
\centering
\subfigure[The degrees of polynomials are $(14,12,12)$]{\includegraphics[height=5cm]{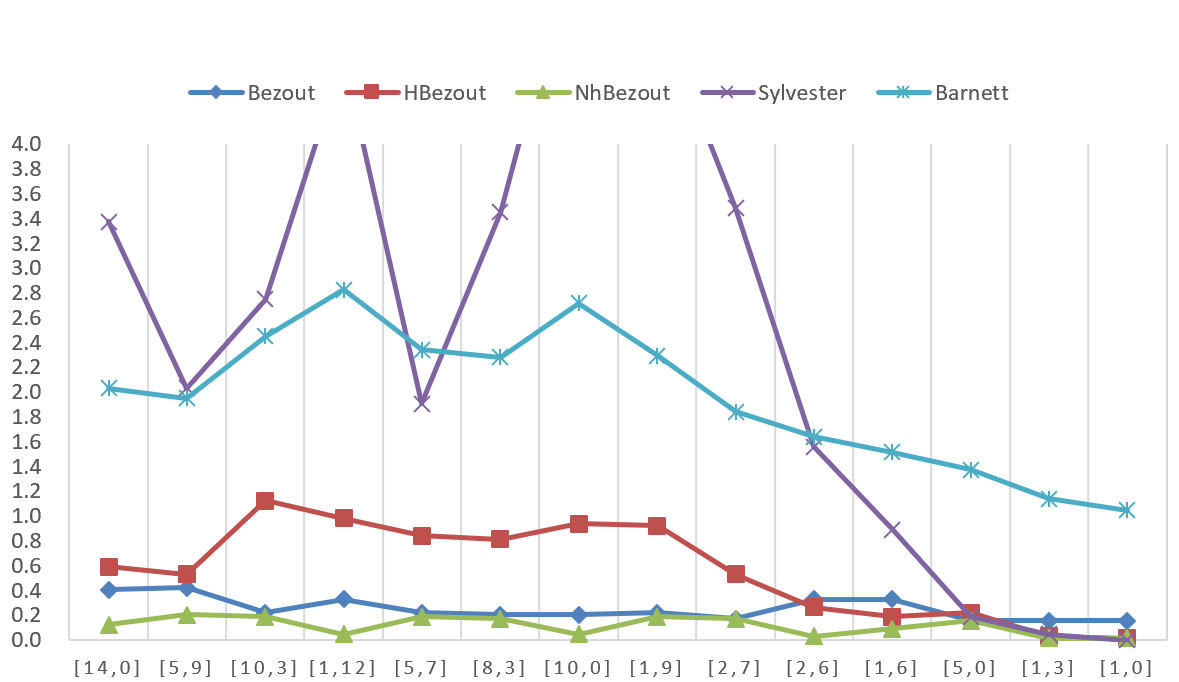}}
\caption{Time cost for computing $S_{\delta}$'s for two polynomials sets  by the listed formulas (where the hybrid B\'ezout type and non-homogeneous B\'ezout type are abbreviated as HBezout and NhBezout, respectively and the horizontal axis stands for the time cost counted with seconds.)  }\label{fig:comparison}
\end{figure}

Figure \ref{fig:comparison} illustrates the cost for two polynomial sets as $\delta$ changes. The degrees of the involving polynomials are $(15, 12, 9)$ and $(14, 12, 12)$ while
the number of parameters are both 2.
Considering the total numbers of possible $\delta$'s are 120 and 136 respectively, in the two examples, it is impractical to list all of them. Thus we select 14 $\delta$'s for each case. In Fig. \ref{fig:comparison} below, the time changes are described by broken lines with different colors. It is seen that the three B\'ezout-type formulas behave better than the other two (i.e., the Sylvester type and Barnett type). Moreover, the non-homogeneous B\'ezout type shows the least time consumption.

To get a better understanding on the time efficiency of the three B\'ezout type formulas, we make a further profiling on them. With some analysis on the program, we identify two operations that cover most of the running time,
which are matrix generation and determinant calculation. In Table \ref{tab:experiments},
we show the time cost for each operation with 10 test examples.
The total time cost listed in the table is the sum of time cost for all possible $\delta$'s and the numbers of involved parameters are all 2. It is seen that in most cases, the non-homogeneous B\'ezout formula dominates all the three formulas while the hybrid B\'ezout behaves worst. However, after a closer look, it is found that the time for generating the hybrid B\'ezout matrix takes almost no time compared with other two formulas. The calculation of determinants takes up almost all the time. Then it naturally leads to a question: is there an efficient method for computing the determinant of a hybrid B\'ezout matrix with its structure to be fully exploited?
This
topic is an interesting topic that needs to be further studied.

\begin{table}[H]
\centering
\caption{The profiling for time cost (in seconds) charged by  two key steps in the computation of $S_{\delta}$'s with three B\'ezout-type subresultant formulas (where $T$ is the total time cost, $M$ is the time cost for generating the subresultant matrix, and $D$ is for calculating the determinant)}\label{tab:experiments}
{\setlength{\tabcolsep}{1.5mm}
\begin{tabular}{c|rrr|rrr|rrr}
\hline
{$d= $}&
\multicolumn{3}{c}~~~B\'ezout\quad~~~~~~
& \multicolumn{3}{c}\!\!\!\!Nonhomogenous\,B\'ezout
& \multicolumn{3}{c}~Hybrid B\'ezout~~~~~~ \\
\cline{2-10}
$\deg F$ & $T$~~ & $M$~~  & $D$~~ & $T$~~ & $M$~~ & $D$~~ &$T$~~ & $M$~~&$D$~~               \\
\cline{1-10}
(12,11,10) & 11.300  & 6.155 & 5.097 & 7.509  & 2.237  & 5.240  &40.412  &0.000     &  40.334        \\
\cline{1-10}
(12,11,10) &11.193 & 6.886 & 4.327 & 7.155  & 2.278  & 4.861 &40.876  &0.000     &  40.719        \\
\cline{1-10}
(13,10,10) &7.934  & 4.764 & 3.155 & 5.547  & 2.128  & 3.387 &22.423  &0.000     &   22.392        \\
\cline{1-10}
(13,10,10) &7.151  & 4.027 & 3.124 & 5.350   & 2.526  & 2.824 &21.346  &0.000     &   21.299          \\
\cline{1-10}
(16,12,10)&33.030   & 23.890 & 9.125 & 26.797 & 7.780   & 19.017&120.701 &0.000     &   120.544   \\
\cline{1-10}
(16,12,10) &32.167  & 23.246& 8.906 & 25.510  & 6.362  & 19.116&119.450  &0.016 &   119.263   \\
\cline{1-10}
(13,12,12) &12.418  & 8.781 & 3.622 & 4.750   & 2.031  & 2.704 &48.396  &0.000     &   48.302   \\
\cline{1-10}
(13,12,12)&11.316  & 8.397 & 2.919 & 4.200    & 1.563  & 2.637 &47.029  &0.000     &   46.951   \\
\cline{1-10}
(14,10,5) &9.036   & 5.860  & 3.161 & 7.815  & 1.686  & 6.129 &17.045  &0.000     &   16.998   \\
\cline{1-10}
(14,10,5) &6.020    & 3.548 & 2.472 & 6.298  & 1.237  & 5.061 &15.727  &0.000     &   15.649   \\
 \cline{1-10}
\end{tabular}}
\end{table}

\medskip\noindent\textbf{Acknowledgements.} The authors' work was
supported by National Natural Science Foundation of China (Grant Nos. 12261010 and 11801101), Natural Science Foundation of Guangxi (Grant No. AD18126010) and  the Natural Science Cultivation Project of Guangxi Minzu University  (Grant No. 2022MDKJ001).


\begin{thebibliography}{10}
\bibitem{arnon1984}
Dennis~S Arnon, George~E Collins, and Scott McCallum.
\newblock Cylindrical algebraic decomposition I: The basic algorithm.
\newblock {\em SIAM Journal on Computing}, 13(4):865--877, 1984.

\bibitem{asadi2022subresultant}
Mohammadali Asadi, Alexander Brandt, David~J. Jeffrey, and Marc~Moreno Maza.
\newblock Subresultant chains using B{\'e}zout matrices.
\newblock In Fran{\c{c}}ois Boulier, Matthew England, Timur~M. Sadykov, and
  Evgenii~V. Vorozhtsov, editors, {\em Computer Algebra in Scientific
  Computing}, pages 29--50, Cham, 2022. Springer International Publishing.

\bibitem{barnett1971greatest}
Stephen Barnett.
\newblock Greatest common divisor of several polynomials.
\newblock In {\em Mathematical proceedings of the Cambridge philosophical
  society}, volume~70, pages 263--268. Cambridge University Press, 1971.

\bibitem{barnett1983}
Stephen Barnett.
\newblock {\em Polynomials and linear control systems}.
\newblock Marcel Dekker, Inc., 1983.

\bibitem{bostan2017}
Alin Bostan, Carlos D'Andrea, Teresa Krick, Agn{\`e}s Szanto, and Marcelo
  Valdettaro.
\newblock Subresultants in multiple roots: an extremal case.
\newblock {\em Linear Algebra and its Applications}, 529:185--198, 2017.

\bibitem{collins1967}
George~E Collins.
\newblock Subresultants and reduced polynomial remainder sequences.
\newblock {\em Journal of the ACM (JACM)}, 14(1):128--142, 1967.

\bibitem{collins1991}
George~E Collins and Hoon Hong.
\newblock Partial cylindrical algebraic decomposition for quantifier
  elimination.
\newblock {\em Journal of Symbolic Computation}, 12(3):299--328, 1991.

\bibitem{cox2021}
David~A Cox and Carlos D'Andrea.
\newblock Subresultants and the shape lemma.
\newblock {\em arXiv preprint arXiv:2112.10306}, 2021.

\bibitem{diaz2002}
Gema~M Diaz-Toca and Laureano Gonzalez-Vega.
\newblock Barnett's theorems about the greatest common divisor of several
  univariate polynomials through Bezout-like matrices.
\newblock {\em Journal of Symbolic Computation}, 34(1):59--81, 2002.

\bibitem{diaz2004various}
Gema~M Diaz-Toca and Laureano Gonzalez-Vega.
\newblock Various new expressions for subresultants and their applications.
\newblock {\em Applicable Algebra in Engineering, Communication and Computing},
  15(3):233--266, 2004.

\bibitem{hy2021}
Hoon Hong and Jing Yang.
\newblock A condition for multiplicity structure of univariate polynomials.
\newblock {\em Journal of Symbolic Computation}, 104:523--538, 2021.

\bibitem{hong2021subresultant}
Hoon Hong and Jing Yang.
\newblock Subresultant of several univariate polynomials.
\newblock {\em arXiv preprint arXiv:2112.15370}, 2021.

\bibitem{houwang2000}
Xiaorong Hou and Dongming Wang.
\newblock Subresultants with the B{\'e}zout matrix.
\newblock In {\em Computer Mathematics}, pages 19--28. World Scientific, 2000.

\bibitem{kapur1994}
Deepak Kapur, Tushar Saxena, and Lu~Yang.
\newblock Algebraic and geometric reasoning using dixon resultants.
\newblock In {\em Proceedings of the international symposium on Symbolic and
  algebraic computation}, pages 99--107, 1994.

\bibitem{lascoux2003}
Alain Lascoux and Piotr Pragacz.
\newblock Double Sylvester sums for subresultants and multi-Schur functions.
\newblock {\em Journal of Symbolic Computation}, 35(6):689--710, 2003.

\bibitem{li2006}
Yong-Bin Li.
\newblock A new approach for constructing subresultants.
\newblock {\em Applied mathematics and computation}, 183(1):471--476, 2006.

\bibitem{sylvester1853}
Sylvester.
\newblock On a theory of syzygetic relations of two rational integral
  functions, comprising an application to the theory of Sturm's functions, and
  that of the greatest algebraic common measure.
\newblock {\em Phil. Trans}, 143:407--548, 1853.

\bibitem{terui2008}
Akira Terui.
\newblock Recursive polynomial remainder sequence and its subresultants.
\newblock {\em Journal of Algebra}, 320(2):633--659, 2008.

\bibitem{wang1998}
Dongming Wang.
\newblock Decomposing polynomial systems into simple systems.
\newblock {\em Journal of Symbolic Computation}, 25(3):295--314, 1998.

\bibitem{wang2000}
Dongming Wang.
\newblock Computing triangular systems and regular systems.
\newblock {\em Journal of Symbolic Computation}, 30(2):221--236, 2000.

\end{thebibliography}
\end{document}